\documentstyle[11pt]{book}
\newcommand{\ch}{{\rm ch}}
\newcommand{\be}{\begin{equation}}
\newcommand{\ee}{\end{equation}}
\newcommand{\bea}{\begin{eqnarray}}
\newcommand{\eea}{\end{eqnarray}}

\newcommand{\Anti}{{\bf Anti}}
\newcommand{\Sym}{{\bf Sym}}

\newcommand{\N}{{\bf N}}

\newcommand{\ov}{\overline}
\def\IR{\relax{\rm I\kern-.18em R}}

\def\IP{\relax{\rm I\kern-.18em P}}
\def\inbar{\vrule height1.5ex width.4pt depth0pt}
\def\IC{\relax\,\hbox{$\inbar\kern-.3em{\rm C}$}}

\begin{document}
\chapter*{ }
\vskip -5truecm
\vskip 15truemm
\begin{flushright} \vspace{-2cm}
{\small MPP-2005-102}\end{flushright}
\vspace{1.0cm}
\centerline {\bf Non-Abelian Brane Worlds:
    The Open String Story} 
\vskip 10truemm
\centerline {\bf Ralph Blumenhagen, Gabriele Honecker, Timo Weigand }  
\vskip 5truemm
\centerline{{Max-Planck-Institut f\"ur Physik, F\"ohringer Ring 6, \\
  80805 M\"unchen, Germany }}


\setcounter{page}{1}

\markboth{Ralph Blumenhagen, Gabriele Honecker, Timo Weigand } 
{Chiral Type I  Compactifications  on Calabi-Yau
Manifolds }  

\begin{quote}
\small{\bf Abstract.} {\it We extend the string model building rules for
the construction of chiral supersymmetric Type I compactifications
on smooth Calabi-Yau manifolds. These models contain
stacks of D9-branes endowed with general stable $U(n)$ bundles on their
world-volume and D5-branes wrapping holomorphic curves on the Calabi-Yau.}

\end{quote}

\begin{quote}
\small{\bf Key words:} {\it String compactifications, D-branes  }
\end{quote}
\vskip 5truemm

\centerline {\sc 1. INTRODUCTION}

Considerable effort has gone into the stringy construction of
semi-realistic intersecting D-brane models during the last five years
(see \cite{Uranga:2003pz,Kiritsis:2003mc,Lust:2004ks,Blumenhagen:2005mu} for reviews and further references.). 
Though people have tried  hard, it is fair to say that so far no completely satisfactory
model has emerged. However, before jumping to conclusions, one should keep
in mind that all the effort has essentially focused on a very tiny subset of concrete models. 
For simplicity one has studied in detail the restricted class of a handful of toroidal 
orbifold spaces with a certain subset of supersymmetric D-branes. 
In the T-dual (mirror symmetric) picture these intersecting D-brane models are
Type IIB orientifolds with magnetized D9-branes. The magnetized
D-branes carry $U(1)$ bundles on their world-volume, and the supersymmetry
condition in flat space is just the abelian MMMS equation \cite{Marino:1999af}.

Heterotic string model building and its recent successes 
\cite{Donagi:2004ia,Braun:2005bw}
demonstrate that more generally it might prove fruitful to consider arbitrary Calabi-Yau spaces equipped with 
vector bundles with structure groups of higher rank. In fact,
we have studied compactifications of both the 
$E_8\times E_8$ \cite{Blumenhagen:2005ga}(see also \cite{Lukas:1999nh,Andreas:2004ja})
 and the 
$SO(32)$ \cite{Blumenhagen:2005pm} heterotic string  endowed
with general $U(n)$ bundles. Note that before, mostly the $E_8\times E_8$
heterotic string with $SU(n)$ bundles was considered. 
Of course, one expects that the general results 
for the $SO(32)$ heterotic string \cite{Blumenhagen:2005pm} can be translated via S-duality to 
compactifications of the Type I string. In this article
we explicitly verify this statement and derive the model
building rules for the Type I compactifications on smooth Calabi-Yau
manifolds  directly from the Type IIB and D-brane perspective. 
Here we focus just on usual $\Omega$ orientifolds, but the
result for more general orientifolds including an $\Omega$-dressing by some
holomorphic involution of the Calabi-Yau can be derived in a similar fashion. \\
\indent The aim of this article  is to make clear that the class of intersecting/magnetized D-brane models
constitutes only a tiny subset 
of the more general class of string vacua obtained by compactifying  general Type IIB orientifolds on smooth Calabi-Yau spaces
and introducing D-branes endowed with general vector bundles with unitary structure groups.
 Concretely, we summarize the model building rules for the construction
of such models. In particular, in Section 2, by dimensional reduction
of the Chern-Simons terms on  D-branes, we derive the 10-form and 6-form tadpole
cancellation conditions. In Section 3 we provide rules  for computing
the chiral as well as non-chiral massless open string spectrum, which due to the Riemann-Roch-Hirzebruch 
theorem is determined by the Euler characteristics of various vector bundles.
In addition to the D9-branes, we also allow for D5-branes wrapping effective 2-cycles
on the Calabi-Yau. These branes carry symplectic gauge groups, which allows
us in Section 4 to compute the K-theory constraints from the vanishing of the
global Witten anomaly on these D5-branes. 
In Section 5 we summarize the main formulas for the cancellation of the various abelian
anomalies. These are used to derive in Section 6 the perturbative expressions for the Fayet-Iliopolous (FI)
terms and in Section 7 the holomorphic gauge kinetic functions. 
These two quantities are given by the same expression
as appearing  in the $\Pi$-stability condition for B-type branes \cite{Douglas:2000ah}.
In Section 8 we summarize the supersymmetry conditions for the magnetized D9-branes and
conclude in Section 9 with some outlook on the next steps to concrete semi-realistic
string model building using the set-up.

\vskip 0.3cm
\centerline {\sc 2. TADPOLE CANCELLATION}
\vskip 0.1cm
We consider compactifications of the Type I string to four space-time
dimensions on a  Calabi-Yau manifold $X$.
 We start with the ambient model, which is 
the Type IIB string divided by the world-sheet parity 
transformation $\Omega:(\sigma,\tau)\to (-\sigma,\tau)$.
As is well known, this induces a tadpole for the Ramond-Ramond (R-R)
10-form, $C_{10}$, and, since  the Calabi-Yau is generically curved,
an induced tadpole for the 6-form $C_{6}$.
Quantitatively, these tadpoles are given by the CS-terms on the
$O9$-plane \cite{Green:1996dd,Blumenhagen:2002wn}
\bea
\label{cs1}
  S^{CS}_{O9}=  -32\,\mu_9\,\int_{\IR^{1,3}\times X} \left(\sum_{n=0}^2
       C_{4n+2}\right)\wedge \sqrt{\hat{\cal L}
      \left({{\cal R}\over 4}\right) },
\eea
where ${\cal R}=-i\ell_s^2 R$ with the string length defined as
$\ell_s=2\pi\sqrt{\alpha'}$. The Hirzebruch genus $\hat{\cal L}$ is
defined as
\bea
    \sqrt{\hat{\cal L}\left({{\cal R}\over 4}\right)}
   =1+{\ell_s^4\over 192\, (2\pi)^2}\, {\rm tr} R^2 + 
     {\ell_s^8\over 73728\, (2\pi)^4}\, \left({\rm tr} R^2\right)^2 -
     {\ell_s^8\over 92160\, (2\pi)^4}\, \left({\rm tr} R^4\right).
\eea
The traces are taken over the fundamental representation
of the Lorentz group $SO(1,9)$.

In order to cancel these tadpoles, one introduces D9-branes endowed
with  holomorphic vector bundles (coherent sheaves) on their 
world-volume.
More concretely, we take stacks of $M_i=N_i\, n_i$ branes and diagonally 
turn on $U(n_i)$ holomorphic vector bundles $V_i$, breaking the observable
gauge group to $\prod_i U(N_i)$.  If the gauge field on such a stack is $F_i$,
then under the action of $\Omega$ this stack is mapped to a different
stack with gauge field $-F_i$. Therefore, we have to introduce 
these stacks in pairs with vector bundles $V_i$ and $V_i^*$ supported
on their world-volume.  

The Chern-Simons action on the D9-branes reads
\bea
\label{cs2}
  S^{CS}_{D9_i}=  2\,  \mu_9\,\int_{\IR^{1,3}\times X} \left(\sum_{n=0}^2
       C_{4n+2}\right)\wedge \ch(i{\cal F}_i)\wedge \sqrt{\hat{\cal A}
      \left({{\cal R}}\right) } 
\eea
with ${\cal F}=-i\ell_s^2 F$, $\mu_9={1\over (2\pi)^9 \alpha'^5}$ and
\bea
   \ch_k(i{\cal F}_i)&=&{\ell_s^{2k}\over k!\, (2\pi)^k}{\rm Tr}_{M_i}(F_i^k),\\
    \sqrt{\hat{\cal A}\left({{\cal R}}\right)}
   &=&1-{\ell_s^4\over 96\, (2\pi)^2}\, {\rm tr} R^2 + 
     {\ell_s^8\over 18432\, (2\pi)^4}\, \left({\rm tr} R^2\right)^2 + \\
   &&\phantom{aaaaaaaaaaaaaaaaaaaaaaaaa}  
      {\ell_s^8\over 11520\, (2\pi)^4}\, \left({\rm tr} R^4\right). \nonumber
\eea

In addition, we allow for stacks of $2N_a$ $D5$-branes wrapping 
holomorphic 2-cycles, $\Gamma_a$, 
on $X$. The Chern-Simons action on the D5-branes reads
\bea
\label{cs3}
  S^{CS}_{D5_a}=  -\mu_5\,\int_{\IR^{1,3}\times \Gamma_a} 
\left(\sum_{n=0}^1
       C_{4n+2}\right) \wedge  \left(2\, N_a+
     {\ell_s^{4}\over 2\, (2\pi)^2}{\rm Tr}_{SP}(F_a^2)\,
  \right)\wedge {\sqrt{\hat{\cal A}
      \left({\rm T}\Gamma_a\right)}\over 
       \sqrt{\hat{\cal A}
      \left({{\rm N}\Gamma_a}\right) }} 
\eea
with $\mu_5={1\over (2\pi)^5 \alpha'^3}$.
Here T$\Gamma_a$ denotes the tangent bundle and 
N$\Gamma_a$ the  normal bundle of the D5-brane in $X$. 
The gauge group on such a stack of D5-branes is $SP(2N_a)$.
If we also allowed for $2M$ D9-branes with trivial gauge bundle, these
would support an additional $SO(2M)$ gauge factor. For shortness
we do not explicitly include these branes in our formulas, but
this is easily accomplished. 

From the CS terms it is straightforward to derive the tadpole cancellation
condition for $C_{10}$ and $C_6$ 
\bea
\label{tad1}
             \sum_{i=1}^K N_i\, n_i &=& 16, \\
         \sum_{i=1}^K N_i\, \ch_2(V_i)-\sum_{a=1}^L 
               N_a\, \gamma_a  &=& -c_2(T), \nonumber
\eea
where $\gamma_a$ denotes  the Poincare dual 4-form of the 2-cycle $\Gamma_a$.

\vskip 0.5cm
\centerline {\sc 3. MASSLESS SPECTRUM}

The  chiral massless spectrum resulting from open strings stretched between
the different stacks of D9 and D5-branes is determined by the
respective Euler characteristics  
\bea
            \chi(W)=\sum_{r=0}^3 (-1)^r {\rm dim} \, H^r(X,W)= 
\int_X \left( \ch_3(W)+{1\over 12}\, c_1(W)\, c_2(T) \right)
\eea
listed in Table 1.

\begin{table}[htb]
\renewcommand{\arraystretch}{1.5}
\begin{center}
\begin{tabular}{|c||c|}
\hline
\hline
reps. & $\prod_{i=1}^K SU(N_i)\times U(1)_i \times \prod_{a=1}^L SP(2N_a)$   \\
\hline \hline
$(\Sym_{U(N_i)})_{2(i)}$ & $\chi(\bigwedge^2 V_i)$  \\
$(\Anti_{U(N_i)})_{2(i)}$ & $\chi(\bigotimes^2_s  V_i)$  \\
\hline
$(\N_i,\N_j)_{1(i),1(j)}$ & $\chi(V_i \otimes V_j)$ \\
$(\N_i,\ov \N_j)_{1(i),-1(j)} $ &  $\chi(V_i \otimes V_j^{\ast})$ \\
\hline
$(\N_i,2\N_a)_{1(i)}$ & $\chi(V_i\otimes {\cal O}\vert_{\Gamma_a})$ \\
\hline
\end{tabular}
\caption{\small Chiral massless spectrum. }
\label{Tchiral1}
\end{center}
\end{table}

For the massless D5-brane matter we have described the D5-brane wrapping
the 2-cycle $\Gamma_a$ by the skyscraper sheaf ${\cal O}\vert_{\Gamma_a}$
supported on the 2-cycle $\Gamma_a$.
The Chern classes of this sheaf are
$\ch({\cal O}\vert_{\Gamma_a})=(0,0,-\gamma_a,0)$
implying 
\bea
  \chi( V_i \otimes{\cal O}\vert_{\Gamma_a})=-\int_{\Gamma_a} c_1(V_i).
\eea

The non-chiral massless spectrum can be determined from the
respective cohomology groups $H^*(X,V\otimes W^*)$ or more generally,
if non locally free sheaves are involved, from the extensions
Ext$_X(V,W)$. In addition, there exists non-chiral
adjoint matter counted by $H^1(X,V_i\otimes V_i^*)$ for the
D9-branes and anti-symmetric matter counted by  $H^1(\Gamma_a,{\cal O})$ plus
 $H^0(\Gamma_a,{\rm N}\Gamma_a)$ for the D5-branes \cite{Katz:2002gh}. 
Here N$\Gamma_a$ denotes
the normal bundle of the 2-cycle $\Gamma_a$ in $X$.

One can show that for the chiral matter in Table 1 the non-abelian
gauge anomalies in four dimensions precisely cancel if 
the tadpole cancellation conditions (\ref{tad1}) are satisfied.   

\vskip 0.5cm
\centerline {\sc 4. K-THEORY CONSTRAINTS}

It is well known that in intersecting D-brane models, besides the
R-R tadpole cancellation condition additional torsion constraints arise
due to the existence of stable non-BPS branes classified by K-theory \cite{Uranga:2000xp}.
From the effective field theory, these constraints guarantee
the absence of $SP(2N)$ global Witten anomalies on probe branes carrying
such symplectic gauge fields. In our case these are precisely the
D5-branes wrapping 2-cycles of the Calabi-Yau $X$.
Therefore, the cancellation of the Witten anomaly leads to the constraint
\bea
    \sum_{i=1}^K N_i \, \chi(V_i\otimes {\cal O}\vert_{\Gamma_a})
        =0\ {\rm mod}\ 2
\eea
for every 2-cycle $\Gamma_a$. Therefore, this condition is 
precisely the condition for the entire vector bundle 
$W=\bigoplus_{i=1}^K N_i\, V_i$
to admit spinors
\bea
        c_1(W)=\sum_{i=1}^K   N_i\, c_1(V_i) =0\ {\rm mod}\ 2.
\eea
Note that for the heterotic string this condition was derived
from the absence of anomalies in  the two-dimensional non-linear
sigma model~\cite{Witten:1985mj,Freed:1986zx}.

\vskip 0.5cm
\centerline {\sc 5. GREEN-SCHWARZ MECHANISM}

Since all these string models naturally  contain abelian gauge groups,
one also has mixed abelian-non-abelian, mixed abelian-gravitational and
cubic abelian anomalies. As usual in string theory these
anomalies do not  cancel directly but only after axionic couplings are
taken into account.
Let us briefly summarize at least for the three mixed anomalies how
the generalized Green-Schwarz mechanism works in this case.
The mixed $U(1)_i-SU(N_j)^2$ anomaly for $i\ne j$ is given by
\bea
\label{mixeda}
    A_{i;jj}&=&N_i\left( \chi(V_i\otimes V_j)+\chi(V_i\otimes V_j^*)\right)
          \nonumber\\
      &=& 2N_i\, \int_X \left[n_j\ch_3(V_i) + c_1(V_i)\, \ch_2(V_j) +
     {n_j\over 12}\, c_1(V_i)\, c_2(T)\right] 
\eea 
The last expression also holds for the case $i=j$, where
also the contribution from the symmetric and antisymmetric matter
and the tadpole constraint have to be taken into account.
\noindent
The mixed $U(1)_i-SP(N_a)^2$ anomaly is
\bea
\label{mixedd}
    A_{i;aa}&=&N_i\,  \chi(V_i\otimes {\cal O}\vert_{\Gamma_a} )=
    -N_i \int_X c_1(V_i)\wedge \gamma_a.
\eea 

\noindent
For the mixed $U(1)_i-G^2$ anomaly one finds
\bea
\label{mixedb}
    A_{i;GG}&=&\sum_{j\ne i} N_i\, N_j\left( \chi(V_i\otimes
      V_j)+\chi(V_i\otimes V_j^*)\right) +
  \sum_a 2N_i\, N_a \, \chi({\cal O}\vert_{\Gamma_a} \otimes V_i^{\ast}) 
           +\nonumber \\ && N_i\, (N_i-1)\, 
      \chi(\hbox{$\bigotimes$}^2_s  V_i) +
     N_i\, (N_i+1)\, \chi(\hbox{$\bigwedge$}^2  V_i) \\
      &=& N_i\, \int_X \left[24\, \ch_3(V_i) + 
     {1\over 2}\, c_1(V_i)\, c_2(T)\right]. \nonumber 
\eea 

These anomalies have to be canceled by axionic Green-Schwarz
couplings arising from the dimensional reduction of the 
three kinds of CS-terms (\ref{cs1},\ref{cs2},\ref{cs3}). 
We expand the relevant two and six-forms as
\bea
   C_2=C_0^{(2)} + \ell_s^2 \sum_{k=1}^{h_{11}} C_k^{(0)}\, \omega_k,\quad
   C_6=\ell_s^6\, C_0^{(0)}\, {\rm vol}_6 + \ell_s^4 
       \sum_{k=1}^{h_{11}} C_k^{(2)}\, \hat\omega_k, 
\eea
where $ \omega_k$ and $\widehat\omega_k$ are a normalized basis
of 2- and 4-cycles on $X$ with 
\mbox{$\int  \omega_k\wedge  \widehat\omega_l=\delta_{kl}$}.
The four-dimensional 2-forms  $C_0^{(2)},C_k^{(2)}$ are Hodge dual
to the four-dimensional scalars  $C_0^{(0)},C_k^{(0)}$.
 
By dimensional reduction we obtain the following axionic mass terms
in four-dimensions
\bea
\label{mass}
    && M_0={1\over 6\, (2\pi)^5\alpha'} \sum_i  
   N_i \int_{\IR^{1,3}}  C_0^{(2)}\wedge f_i \ \int_X \left[
         {\rm Tr}_{n_i} \ov F^3_i -{1\over 16}\, {\rm Tr}_{n_i} \ov F_i\wedge
           {\rm tr} \ov R^2\right], \nonumber\\
    && M_k={1\over (2\pi)^2\, \alpha'} \sum_i  
   N_i \int_{\IR^{1,3}}  C_k^{(2)}\wedge f_i \ 
         \left[{\rm Tr}_{n_i} \ov F_i\right]_k,
\eea
where $f_i$ denotes the field strength of the $U(1)_i$ observable
gauge group and $\ov F_i$ the field strength of the internal
gauge field.
The traces Tr$_{n_i}$ are over the fundamental representation of the
structure group  $U(n_i)$. 
We have expanded
\bea
{\rm Tr} \ov F_i=(2\pi)\sum_k \left[{\rm Tr} \ov F_i\right]_k\, 
          \omega_k.
\eea
Similarly one obtains the vertex couplings for $C_0^{(0)}$
\bea
\label{vertex0}
   V_0={1 \over  2\, (2\pi)} \sum_i  
   n_i \int_{\IR^{1,3}}  
        C_0^{(0)}\wedge  {\rm Tr}_{N_i}  F^2_i -
   {1\over 4\,  (2\pi)}  \int_{\IR^{1,3}}  
        C_0^{(0)}\wedge  {\rm tr}   R^2 
\eea
and for $C_k^{(0)}$
\bea
\label{vertexk}
 V_k&=&{1 \over  4\, (2\pi)} \sum_i  
      \int_{\IR^{1,3}}  
        C_k^{(0)}\wedge  {\rm Tr}_{N_i}  F^2_i \ 
         \left[ {\rm Tr}_{n_i} \ov F^2_i-{n_i\over 48} 
         {\rm tr} \ov R^2 \right]_k -\nonumber \\
   && {1 \over  4\, (2\pi)} \sum_a  
      \int_{\IR^{1,3}}  
        C_k^{(0)}\wedge  {\rm Tr}_{SP(2N_a)}  F^2_a \ 
         \left[ \gamma_a \right]_k- \\
&& {1 \over 768\,  (2\pi)}  \int_{\IR^{1,3}}  
        C_k^{(0)}\wedge  {\rm tr}  R^2 \ 
         \left[  {\rm tr} \ov R^2 \right]_k. \nonumber 
\eea

Here we have expanded
\bea
{\rm Tr} \ov F^2_i=(2\pi)^2\sum_k \left[{\rm Tr} \ov F^2_i\right]_k\, 
          \widehat\omega_k,\quad\quad 
\gamma_a= \sum_k [\gamma_a]_k\, \widehat\omega_k
\eea
and similarly for the internal curvature.
Note that in the derivation of these vertex couplings also the
D5-branes gave a contribution and that the tadpole cancellation
conditions had to be used to bring the expression to its final
form (\ref{vertexk}).
Now we can combine the axionic mass and vertex couplings to
provide counter terms for the triangle anomalies.
Indeed, adding up all these graphs  yields  precisely an expression of the form of the
mixed anomalies (\ref{mixeda},\ref{mixedd},\ref{mixedb}).

As usual the anomalous $U(1)$ gauge fields become massive 
via the Green-Schwarz couplings, where the longitudinal 
polarisations  are given by some of the massive axionic fields.


\vskip 0.5cm
\centerline {\sc 6.  FAYET-ILIOPOLOUS TERMS}

From the general analysis of four-dimensional ${\cal N}=1$ supergravity 
it is well-known that the 
coefficients $\xi_i$ of the FI-terms can be derived from 
the K\"ahler potential $\cal K$ via the relation
\bea
\label{FIterms}
   {\xi_i\over g_i^2}=   {\partial {\cal K}\over \partial V_i}
\biggr\vert_{V=0},
\eea
where the gauge invariant K\"ahler potential relevant for our type of 
construction reads
\bea
{\cal K} &=&{M^2_{pl}\over 8\pi} \Biggl[   
       -\ln\Biggl(S+S^*-\sum_x Q^i_0\, V_i\Biggr)-\ln\Biggl(-\sum_{k,l,m=1}^{h_{11}}
{d_{klm}\over 6} 
\biggl(  T_k+T_k^*-\sum_i Q^i_k\, V_i\biggr) \nonumber \\
&& \phantom{aaaaaaa} \biggl(  T_l+T_l^*-\sum_i Q^i_l\, V_i\biggr)
\biggl(  T_m+T_m^*-\sum_i Q^i_m\, V_i\biggr) \Biggr) \Biggr].
\eea
The charges $ Q^i_k$ are
defined via 
\bea
                 S_{mass}=\sum_{i=1}^{K}  \sum_{k=0}^{h_{11}} 
            {Q^i_k\over 2\pi\alpha'}
                \int_{\IR_{1,3}} f_i \wedge C^{(2)}_k 
\eea
and  can easily be extracted from the mass terms (\ref{mass}).

This results in the FI-terms 
\bea
\label{fiterms}
   {\xi_i\over g_i^2}\simeq  {1\over 2} \int_X J\wedge J\wedge {\rm Tr}_{n_i}
          \ov F_i -{(2\pi\alpha')^2\over 3!}\, \int_X
   \left[{\rm Tr}_{n_i} \ov F^3_i -{1\over 16}\, {\rm Tr}_{n_i} \ov F_i\wedge
           {\rm tr} \ov R^2\right].
\eea
Since they depend on the K\"ahler moduli, though  exact in sigma model
perturbation theory, one expects these expressions
to be corrected by world-sheet instanton contributions.
Supersymmetry implies that the D-terms have to vanish, which
for zero VEVs for charged matter fields means that all
FI-terms have to vanish. Note that setting (\ref{fiterms}) to zero
is nothing else than the non-abelian generalization of the MMMS
equation also including curvature terms.
 
The FI-term can be written as the imaginary part of a central charge
\bea
\label{fiterms2}
   {\xi_i\over g_i^2}\simeq  
    {\rm Im}\left(  \int_{X} {\rm Tr}_{n_i}
\left[ e^{-i\varphi} \, e^{2\pi\alpha' F -iJ}\,
         \sqrt{\hat A(X)} \right]\right) 
\eea
with $\varphi=\pi/2$.
This is precisely the perturbative part of 
the expression appearing in the $\Pi$-stability condition of \cite{Douglas:2000ah}.
In the case of an $\Omega\sigma$ orientifold with O7- and induced O3-planes, for the 
introduced pairs of $D9-\ov{D9}$ branes one would get a similar result with $\varphi=0$.

\vskip 0.5cm
\centerline {\sc 6.  GAUGE KINETIC FUNCTIONS}

Let us now give the expressions for  the gauge kinetic functions.
The holomorphic gauge kinetic function ${\rm f}_i$ appears in the four dimensional effective field 
theory as 
\bea
       {\cal L}_{YM}={1\over 4}\, {\rm Re} ({\rm f}_i)\, F_i\wedge\star F_i + {1 \over 4}\,
         {\rm Im} ({\rm f}_i)\, F_i\wedge  F_i. 
\eea
With the definition of the complexified dilaton and  K\"ahler moduli
\bea
    S={1\over 2\pi}\left[ e^{-\phi_{10}} {{\rm Vol}({\cal M}) \over
             \ell_s^6 } + {i}\, C^{(0)}_0 \right],
\qquad\qquad
    T_k={1\over 2\pi}\left[ - e^{-\phi_{10}} \alpha_k + {i} C^{(0)}_k \right],
\eea
the gauge kinetic functions can be deduced from their imaginary parts in
the vertex couplings (\ref{vertex0},\ref{vertexk})
\bea
\label{ggg}
{\rm f}_{SU(N_i)} &=& 2 n_i\, S +\sum_{k=1}^{h_{11}} T_k\,
\left[{\rm Tr}_{n_i} \ov F_i^2-\frac{n_i}{48} {\rm tr} \ov R^2\right]_k .
\eea

The real part of the holomorphic gauge kinetic function ${\rm f}_i$
can be cast into the form
\bea
\label{gal}
{\rm Re}({\rm f}_{i}) = {1\over \pi\ell_s^6 g_s}\left[ {n_i\over 3!}\, 
 \int_{X} J \wedge J \wedge J -
{(2\pi\alpha')^2\over 2} \int_{X} J \wedge 
       \left( {\rm Tr}_{n_i} \ov {F}_i^2  - {n_i \over 48}  {\rm tr}{\ov R}^2 \right)
\right]
\eea
and further be written as
\bea
  {\rm Re}({\rm f}_{i}) \simeq  
    {\rm Re}\left(  \int_{X} {\rm Tr}_{n_i}
\left[  e^{-i\varphi} \, e^{2\pi\alpha' F -iJ}\,
         \sqrt{\hat A(X)} \right]\right) 
\eea
with $\varphi=\pi/2$. 
For the D5-branes the gauge couplings are given by
\bea
   {\rm Re}({\rm f}_{a}) = {1\over 2\pi\ell_s^2 g_s} \int_{\Gamma_a} J. 
\eea

\vskip 0.5cm
\centerline {\sc 7.  SUPERSYMMETRY}

For the mostly studied case of choosing just  $U(1)$ bundles the supersymmetry condition
was simply the vanishing of the FI-terms (\ref{fiterms}). 
In \cite{Blumenhagen:2005pm} arguments were presented 
that the non-integrated supersymmetry condition in the large
radius limit reads 
\bea 
\label{SUSYloc}
   \left[ {\rm Im}\left(  e^{-i\varphi} \, e^{2\pi\alpha' F -iJ}\,
         \sqrt{\hat A(X)} \right)\right]_{\rm top} =0.
\eea
The notion of stability relevant for (\ref{SUSYloc})  has been 
analysed in \cite{Enger:2003ue} and  called $\pi$-stability 
(to stress that it is only the perturbative part of $\Pi$-stability \cite{Douglas:2000ah}). 
In particular, the authors have shown that in the large radius limit (\ref{SUSYloc}) has a unique solution 
precisely if the bundle is stable with respect to the deformed slope 
\bea
\label{cite}
\pi({V}) = -{\rm Arg} \left( \int_{X} {\rm Tr}_{n_i}
\left[ e^{2\pi\alpha' F -iJ}\,
         \sqrt{\hat A(X)}\right] \right),
\eea
i.e. the phase of the central charge.
For supersymmetric configurations, we 
need to ensure that all objects are BPS with respect to the same supersymmetry 
algebra. This is guaranteed by the integrability condition given by the
vanishing of the FI-terms (\ref{fiterms2}) 
(with $\varphi =\pi/2$ in our case). For each stack of D9-branes
one obtains one constraint on the K\"ahler moduli, so that a certain number 
of them, together with
their axionic partners, are  frozen (if we set all VEVs of charged fields to zero).

Not very much is known about $\pi$-stable bundles, but it has been
shown that  at large radius $\mu$-stability implies $\pi$-stability \cite{Enger:2003ue}, so that one can use
the well studied class of $\mu$-stable bundles for concrete model building.

\vskip 0.5cm
\centerline {\sc 8.  TOWARDS STRING MODEL BUILDING}

We have collected the main large radius model building ingredients and consistency 
conditions for the construction of chiral supersymmetric Type I models
with D9-branes endowed with stable unitary bundles as well as D5-branes wrapping effective cycles. 
For the gauge kinetic functions and the FI-terms 
there will be further world-sheet instanton corrections, so
that our analysis in only correct in the perturbative regime. 

The next step is to do concrete model building and to look for Standard-like  
or GUT like models.
So far this program has only been carried out for simple toroidal orbifold spaces
with D9-branes endowed with just $U(1)$ bundles. 
Since one needs to have certain control over stable bundles, a good starting
point is the spectral cover construction of $\mu$-stable $SU(n)$ bundles on elliptically
fibered Calabi-Yau spaces \cite{Friedman:1997yq}. 

One way to define stable $U(n)$ bundles is via twisting an $SU(n)$ bundle with 
a line bundle on $X$. 
One  starts with a stable bundle $SU(n)$ bundle $V$ as it
arises  from the spectral cover construction of Friedman, Morgan
and Witten \cite{Friedman:1997yq,Curio:1998vu,Andreas:2004ja}.
In addition  we take an arbitrary line bundle ${\cal Q}$ on $X$. 
Then one  can define the twisted bundle $V_{{\cal Q}}=V\otimes {\cal Q}$,
which has non-vanishing first Chern class  unless ${\cal Q}$ is trivial.
A bundle $V$ is $\mu$-stable if
and only if $V\otimes {\cal Q}$ is stable for every line bundle ${\cal Q}$ \cite{Mumford,Takemoto}.
Therefore, all these twisted $U(n)$ bundles are $\mu$-stable
if the $SU(n)$ bundles are.
A couple of semi-realistic models have been constructed in the S-dual heterotic
framework in \cite{bhwnew}.
A statistical analysis similar to \cite{Blumenhagen:2004xx} would be an 
interesting task to perform.

{\bf Acknowledgement}: R.B. would like to thank the organizers
  of  the {\it Southeastern European Workshop
Challenges Beyond the Standard Model}, 19-23 May 2005,
Vrnjacka Banja, Serbia for the invitation.

\renewcommand {\bibname} {\normalsize \sc References}

{}


\begin{thebibliography}{99}

\bibitem{Uranga:2003pz}
A.~M.~Uranga,
Class.\ Quant.\ Grav.\  {\bf 20}, S373 (2003)
[arXiv:hep-th/0301032].

\bibitem{Kiritsis:2003mc}
  E.~Kiritsis,
  Fortsch.\ Phys.\  {\bf 52} (2004) 200
  [arXiv:hep-th/0310001].

\bibitem{Lust:2004ks}
D.~L{\"u}st,
Class.\ Quant.\ Grav.\  {\bf 21}, S1399 (2004)
[arXiv:hep-th/0401156].

\bibitem{Blumenhagen:2005mu}
R.~Blumenhagen, M.~Cvetic, P.~Langacker and G.~Shiu,
arXiv:hep-th/0502005.

\bibitem{Marino:1999af}
M.~Marino, R.~Minasian, G.~W.~Moore and A.~Strominger,
JHEP {\bf 0001}, 005 (2000)
[arXiv:hep-th/9911206].

\bibitem{Donagi:2004ia}
R.~Donagi, Y.~H.~He, B.~A.~Ovrut and R.~Reinbacher,
JHEP {\bf 0412}, 054 (2004)
[arXiv:hep-th/0405014].

\bibitem{Braun:2005bw}
V.~Braun, Y.~H.~He, B.~A.~Ovrut and T.~Pantev,
JHEP {\bf 0506}, 039 (2005)
[arXiv:hep-th/0502155];
Phys.\ Lett.\ B {\bf 618}, 252 (2005)
[arXiv:hep-th/0501070].

\bibitem{Blumenhagen:2005ga}
R.~Blumenhagen, G.~Honecker and T.~Weigand,
JHEP {\bf 0506}, 020 (2005)
[arXiv:hep-th/0504232].


\bibitem{Lukas:1999nh}
A.~Lukas and K.~S.~Stelle,
JHEP {\bf 0001}, 010 (2000)
[arXiv:hep-th/9911156].

\bibitem{Andreas:2004ja}
  B.~Andreas and D.~Hernandez Ruiperez,
  arXiv:hep-th/0410170.


\bibitem{Blumenhagen:2005pm}
R.~Blumenhagen, G.~Honecker and T.~Weigand,
JHEP {\bf 0508}, 009 (2005)
[arXiv:hep-th/0507041];



\bibitem{Douglas:2000ah}
M.~R.~Douglas, B.~Fiol and C.~Romelsberger,
arXiv:hep-th/0002037.

\bibitem{Green:1996dd}
M.~B.~Green, J.~A.~Harvey and G.~W.~Moore,
Class.\ Quant.\ Grav.\  {\bf 14}, 47 (1997)
[arXiv:hep-th/9605033];
  R.~Minasian and G.~W.~Moore,
  JHEP {\bf 9711} (1997) 002
  [arXiv:hep-th/9710230];
J.~F.~Morales, C.~A.~Scrucca and M.~Serone,
Nucl.\ Phys.\ B {\bf 552}, 291 (1999)
[arXiv:hep-th/9812071];
B.~J.~Stefanski,
Nucl.\ Phys.\ B {\bf 548} (1999) 275
[arXiv:hep-th/9812088].

\bibitem{Blumenhagen:2002wn}
R.~Blumenhagen, V.~Braun, B.~K{\"o}rs and D.~L{\"u}st,
JHEP {\bf 0207}, 026 (2002)
[arXiv:hep-th/0206038];
arXiv:hep-th/0210083.



\bibitem{Uranga:2000xp}
A.~M.~Uranga,
Nucl.\ Phys.\ B {\bf 598}, 225 (2001)
[arXiv:hep-th/0011048].




\bibitem{Katz:2002gh}
S.~Katz and E.~Sharpe,
Adv.\ Theor.\ Math.\ Phys.\  {\bf 6}, 979 (2003)
[arXiv:hep-th/0208104].


\bibitem{Witten:1985mj}
  E.~Witten,
  ``Global Anomalies In String Theory,''
Print-85-0620 (PRINCETON)
{\it in Proc. of Argonne Symp. on Geometry, Anomalies and Topology, Argonne, IL, Mar 28-30, 1985}

\bibitem{Freed:1986zx}
  D.~S.~Freed,
  ``Determinants, Torsion, And Strings,''
  Commun.\ Math.\ Phys.\  {\bf 107} (1986) 483.



\bibitem{Enger:2003ue}
H.~Enger and C.~A.~Lutken,
Nucl.\ Phys.\ B {\bf 695}, 73 (2004)
[arXiv:hep-th/0312254].


\bibitem{Friedman:1997yq}
R.~Friedman, J.~Morgan and E.~Witten,
Commun.\ Math.\ Phys.\  {\bf 187}, 679 (1997)
[arXiv:hep-th/9701162].

\bibitem{Curio:1998vu}
  G.~Curio,
  ``Chiral matter and transitions in heterotic string models,''
  Phys.\ Lett.\ B {\bf 435} (1998) 39
  [arXiv:hep-th/9803224].


\bibitem{Mumford}
D.~Mumford,
``Geometric Invariant Theory,''
Spinger Verlag 1965.

\bibitem{Takemoto}
F.~Takemoto,
``Stable Vector Bundles on Algebraic Surfaces,''
 Nagoya Math. J. 47(1972) 29.


\bibitem{bhwnew}
R.~Blumenhagen, G.~Honecker and T.~Weigand,
``Non-Abelian Brane Worlds: The Heterotic Story,''
[arXiv:hep-th/0510049].

\bibitem{Blumenhagen:2004xx}
R.~Blumenhagen, F.~Gmeiner, G.~Honecker, D.~L{\"u}st and T.~Weigand,
Nucl.\ Phys.\ B {\bf 713}, 83 (2005)
[arXiv:hep-th/0411173].

\end{thebibliography}
\end{document}